\documentclass[%
 reprint,
superscriptaddress,
amsmath,amssymb,
aps,
prl,
longbibliography,
]{revtex4-2}

\usepackage{xcolor}
\usepackage{graphicx}
\usepackage{dcolumn}
\usepackage{bm}

\begin{document}

\preprint{APS/123-QED}

\title{Reversibility and Multiplicity in Liquid-Vapor Phase Change}

\author{Luiz Eduardo Czelusniak}
\email{luiz.czelusniak@usp.br}
\affiliation{Department of Mechanical Engineering, São Carlos School of Engineering, University of São Paulo (USP), São Carlos, 13566-590, São Paulo, Brazil}

\author{Alexander J. Wagner}
\email{alexander.wagner@ndsu.edu}
\affiliation{Department of Physics, North Dakota State University (NDSU), Fargo, ND 58102, United States}

\author{Luben Cabezas Gómez}
\email{lubencg@sc.usp.br}
\affiliation{Department of Mechanical Engineering, São Carlos School of Engineering, University of São Paulo (USP), São Carlos, 13566-590, São Paulo, Brazil}

\date{\today}

\begin{abstract}
Using diffuse-interface theory, we uncover unexpected phase-change phenomena that are absent from widely used models of evaporation and condensation. In the inviscid limit, the same adjacent thermodynamic states admit stationary evaporation and condensation with equal and opposite mass fluxes, even when the vapor pressure lies below saturation. Viscosity breaks this reversibility but gives rise to multiple stationary condensation rates for the same vapor state. Lattice Boltzmann simulations dynamically recover these states, showing that these phenomena are dynamically realized by the diffuse-interface equations rather than arising solely as mathematical solutions of the reduced interfacial relations. Our results reveal previously unexplored behavior in a well-established theory of liquid--vapor interfaces.
\end{abstract}

\maketitle





\emph{Introduction---}
Quantitative models of evaporation and condensation are central to the
description of liquid--vapor phase change, providing the basis for
analytical predictions~\cite{scriven1995dynamics}, the interpretation of experiments~\cite{persad2016expressions}, and numerical
simulations of multiphase flows~\cite{borah2025enhancing}.
In many interface-resolved numerical methods, including volume-of-fluid~\cite{kunkelmann2009cfd,darshan2025numerical,borah2025enhancing},
phase-field~\cite{sun2019numerical,ronsin2021phase,latifiyan2022numerical,yang2026understanding}, and color-gradient lattice Boltzmann approaches~\cite{nath2025reaction}, phase change is not determined by
the multiphase formulation alone and requires an additional relation
that determines the interfacial mass flux from the conditions adjacent
to the interface.
Such relations range from interfacial energy-balance models based on local thermodynamic equilibrium~\cite{sun2012development} to phenomenological models~\cite{lee1980pressure} and kinetic-theory expressions such as Hertz--Knudsen--Schrage~\cite{hertz1882ueber,knudsen1915maximale,schrage1953theoretical}.

A fundamentally different route is offered by descriptions in which
phase change is already contained in the underlying dynamics, as in
molecular dynamics and single-component diffuse-interface theory~\cite{van1979thermodynamic,korteweg1901forme}.
Rather than introducing a separate phase-change model, such descriptions
offer the possibility of deriving predictive relations for evaporation
and condensation directly from an established underlying theory.
Diffuse-interface theory is particularly attractive in this respect.
Rooted in continuum descriptions of capillarity and nonuniform fluids,
it provides a well-established framework for liquid--vapor interfaces
and underlies widely used free-energy~\cite{zhang2022improved,zhang2026three} and pseudopotential formulations
for multiphase simulation~\cite{li2015lattice,cai2025numerical}.

Interfacial relations have been derived from diffuse-interface models
in the kinetic-relations literature~\cite{slemrod1983admissibility,
dunn1993shock,bedjaoui2002diffusive,zeiler2015liquid,
lefloch2024riemann}.
More recently, we showed that the stationary diffuse-interface equations
can be used quantitatively to predict evaporation states and mass fluxes
for a specified equation of state~\cite{czelusniak2026analytical}.
Yet, despite the widespread use of diffuse-interface models in
multiphase simulations, their potential as analytical and predictive
tools for evaporation and condensation has received comparatively little
attention in the broader phase-change literature~\cite{chen2024review,lorenzini2025analysis}.
Here, we pursue this route systematically by exploring the structure of
evaporation and condensation predicted by the diffuse-interface
equations.

Our analysis uncovers a striking and counterintuitive structure of stationary evaporation and condensation.
In the absence of viscosity, the same adjacent thermodynamic states
admit both evaporation and condensation with equal and opposite mass
fluxes, revealing an unexpected reversibility between the two processes.
Even more remarkably, both branches occur on the same side of
saturation: stationary condensation can coexist with a vapor pressure
below the saturation pressure, a condition conventionally associated
with evaporation~\cite{qu2025experimental,tinguely2012energy,bao2023impact}.
Viscous dissipation breaks this reversibility, but does not restore a
unique correspondence between thermodynamic state and phase-change rate.
Instead, multiple condensation rates can exist for the same adjacent
vapor state.
Thus, diffuse-interface theory predicts that the adjacent thermodynamic
state need not uniquely determine either the direction or the rate of
stationary phase change, a structure with no direct counterpart in
conventional single-valued phase-change relations such as
Hertz--Knudsen--Schrage.

These unexpected states are not merely formal solutions of the reduced
interfacial relations.
Using the lattice Boltzmann method~\cite{chen1998lattice,wagner2006thermodynamic}, we show that these states are dynamically accessible and that distinct stationary states can be selected by the surrounding compressible-flow dynamics.
They therefore represent dynamically accessible consequences of the
underlying diffuse-interface equations rather than artifacts of the
reduced analysis.
Whether this unexpected structure is specific to the diffuse-interface
description or has a counterpart in molecular and experimental
phase-change dynamics poses a fundamental question for future
investigation.





\begin{figure}
    \centering
    \includegraphics[width=0.5\textwidth]{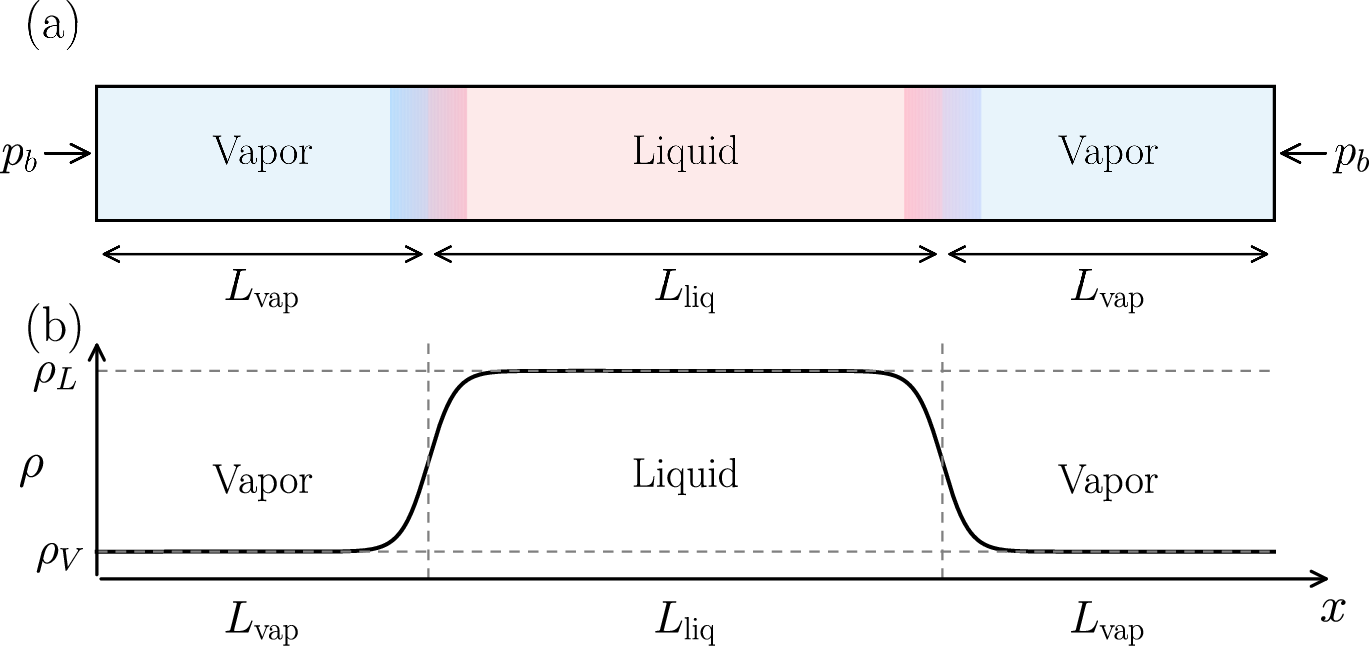}
    \caption{\label{fig:sketch}
    Vapor--liquid--vapor system considered in this work.
    (a) Planar liquid slab surrounded by vapor, with
    $p_{\mathrm b}$ imposed at remote boundaries.
    (b) Equilibrium diffuse-interface profile connecting
    $\rho_{\mathrm{vap}}$ and $\rho_{\mathrm{liq}}$.
    }
\end{figure}

\emph{Diffuse-interface model problem---}
To explore the phase-change behavior predicted by single-component
diffuse-interface theory, we consider the one-dimensional planar
vapor--liquid--vapor configuration shown in Fig.~\ref{fig:sketch}, in
which a liquid slab is surrounded by vapor and the pressure
$p_{\mathrm b}$ can be controlled at remote boundaries. This simplified
setting allows us to investigate how stationary phase-change states are
dynamically selected in response to changes in a remotely imposed
pressure.

The dynamics are governed by the one-dimensional conservation equations
\begin{subequations}
\label{eq:governing_equations}
\begin{equation}
\partial_t \rho + \partial_x(\rho u) = 0,
\label{eq:mass}
\end{equation}
\begin{equation}
\partial_t (\rho u) 
+ \partial_x (\rho u^2) = - \partial_x p 
+ \partial_x \sigma_\nu,
\label{eq:momentum}
\end{equation}
\end{subequations}
where $\rho$ and $u$ denote the density and velocity, respectively, and
$\sigma_\nu$ is the normal viscous stress. In the homogeneous bulk, we
write $\sigma_\nu=2\rho\nu\partial_x u$, where $\nu$ is the kinematic
viscosity. The nonideal pressure $p$
contains both the bulk thermodynamic and capillary contributions and,
for the one-dimensional Korteweg model~\cite{korteweg1901forme} considered here, is given by
\begin{equation}
p
=
p_{\mathrm{EoS}}
+
\frac{\kappa}{2} (\partial_x\rho)^2
-
\kappa \rho
\partial_x^2\rho,
\label{eq:korteweg_pressure}
\end{equation}
where $p_{\mathrm{EOS}}$ is the bulk equation of state and $\kappa$
controls the capillary contribution associated with density gradients.
Together, Eqs.~\eqref{eq:governing_equations} and
\eqref{eq:korteweg_pressure} provide the diffuse-interface dynamics
from which the stationary interfacial relations considered below are
obtained.

\emph{Stationary interfacial relations---}
We now introduce the relations governing stationary phase change across
a diffuse interface. We consider an approximately isothermal interface, for which the temperature variation across the diffuse region is negligible; differences of approximately $0.1\,\mathrm{K}$ or less have been estimated at low evaporation rates~\cite{jafari2020temperature}
(see Sec.~II\,B of the Supplemental Material for further discussion).
Integration of the mass, momentum, and chemical-potential balances
across the diffuse region gives~\cite{slemrod1983admissibility,
dunn1993shock,bedjaoui2002diffusive,zeiler2015liquid} (further details are provided in Sec.~II of the Supplemental
Material),
\begin{subequations}
\label{eq:interfacial_relations}
\begin{equation}
j
=
\rho_{\mathrm{liq}}
\left(
u_{\mathrm{liq}}-u_{\mathrm{int}}
\right)
=
\rho_{\mathrm{vap}}
\left(
u_{\mathrm{vap}}-u_{\mathrm{int}}
\right),
\label{eq:interfacial_mass}
\end{equation}
\begin{equation}
p_{\mathrm{liq}}
+\frac{j^2}{\rho_{\mathrm{liq}}}
=
p_{\mathrm{vap}}
+\frac{j^2}{\rho_{\mathrm{vap}}},
\label{eq:interfacial_momentum}
\end{equation}
\begin{equation}
\mu_{\mathrm{liq}}^{\mathrm{ch}}
+\frac{j^2}{2\rho_{\mathrm{liq}}^2}
-
\int_{\xi_{\mathrm{vap}}}^{\xi_{\mathrm{liq}}}
\frac{\sigma_\nu(x)}{\rho(x)^2}
\frac{d\rho}{dx}\,dx
=
\mu_{\mathrm{vap}}^{\mathrm{ch}}
+\frac{j^2}{2\rho_{\mathrm{vap}}^2},
\label{eq:interfacial_chemical}
\end{equation}
\end{subequations}
Here, $j$ is the mass flux relative to an interface moving with velocity
$u_{\mathrm{int}}$, and $\xi_{\mathrm{vap}}$ and
$\xi_{\mathrm{liq}}$ denote the vapor- and liquid-side boundaries of
the diffuse interface, respectively. The integral in
Eq.~\eqref{eq:interfacial_chemical} represents the viscous contribution
accumulated across the interfacial region. Its evaluation requires a
constitutive description of $\sigma_\nu$ within the interface, which
is specified below when the viscous solutions are considered. The pressure and chemical potential are obtained consistently
from the same bulk Helmholtz free energy; here we use a Landau
free-energy model~\cite{briant2004lattice}.





\emph{Inviscid phase-change solution structure---}
We first consider the inviscid limit of the stationary interfacial
problem. With the viscous contribution in
Eq.~\eqref{eq:interfacial_chemical} set to zero,
Eqs.~\eqref{eq:interfacial_momentum} and
\eqref{eq:interfacial_chemical} depend on the mass flux only through
$j^2$. Consequently, every nonzero solution is accompanied by a second
solution with equal magnitude and opposite sign,
\[
j\longleftrightarrow -j ,
\]
while the adjacent thermodynamic states remain unchanged.

In our previous work~\cite{czelusniak2026analytical}, we solved this
inviscid problem for vapor pressures below saturation,
$p_{\mathrm{vap}}<p_{\mathrm{sat}}$. We retained the evaporation branch
and discarded the opposite-sign solution, since condensation is
conventionally associated with pressures above saturation rather than
below it~\cite{qu2025experimental,tinguely2012energy,bao2023impact}. Here, we reconsider this interpretation. Reversing $j$
changes only the direction of mass transfer, while leaving the adjacent
thermodynamic states unchanged. The second solution therefore
corresponds to stationary condensation, and we retain both branches as
physically admissible solutions of the diffuse-interface equations:
\[
p_{\mathrm{vap}}<p_{\mathrm{sat}}
\longrightarrow \{E,C\},
\qquad |j_E|=|j_C|,
\]
where $E$ and $C$ denote evaporation and condensation, respectively.
Thus, exactly the same adjacent liquid and vapor states admit a
symmetric pair of stationary evaporation and condensation solutions.
As shown below, both branches are dynamically recovered in lattice
Boltzmann simulations, supporting this physical interpretation.
Related reversible phase transitions have been discussed for kinetic
relations with vanishing entropy production~\cite{zeiler2015liquid};
here, the reversible pair follows directly from the nondissipative
diffuse-interface equations~\cite{freistuhler2017phase}.

We next extend the stationary analysis beyond the sub-saturation
regime considered previously. Since condensation is conventionally
associated with an increase in pressure, one might expect a
condensation solution for
$p_{\mathrm{vap}}>p_{\mathrm{sat}}$. Remarkably, we find that no
stationary inviscid phase-change solution exists in this region:
\[
p_{\mathrm{vap}}>p_{\mathrm{sat}}
\longrightarrow \varnothing .
\]
Both members of the reversible pair are therefore confined to
$p_{\mathrm{vap}}<p_{\mathrm{sat}}$ and terminate together at
equilibrium, $p_{\mathrm{vap}}=p_{\mathrm{sat}}$.

The origin of this restriction can be understood from the simultaneous
momentum and chemical-potential balances. These impose a relation
between the pressure and chemical-potential differences across the
interface, while thermodynamic consistency independently requires
$d\mu^{\mathrm{ch}}=dp/\rho$. For
$p_{\mathrm{vap}}>p_{\mathrm{sat}}$, these constraints become
incompatible and admit no real solution for $j^2$. An analytical proof
is given in Sec.~III of the Supplemental Material.

The results below are reported in dimensionless variables, denoted by
the superscript ${}^\ast$. Their definitions are given in Sec.~V\,C of the Supplemental Material.

\begin{figure}[t]
    \centering
    \includegraphics[width=\columnwidth]{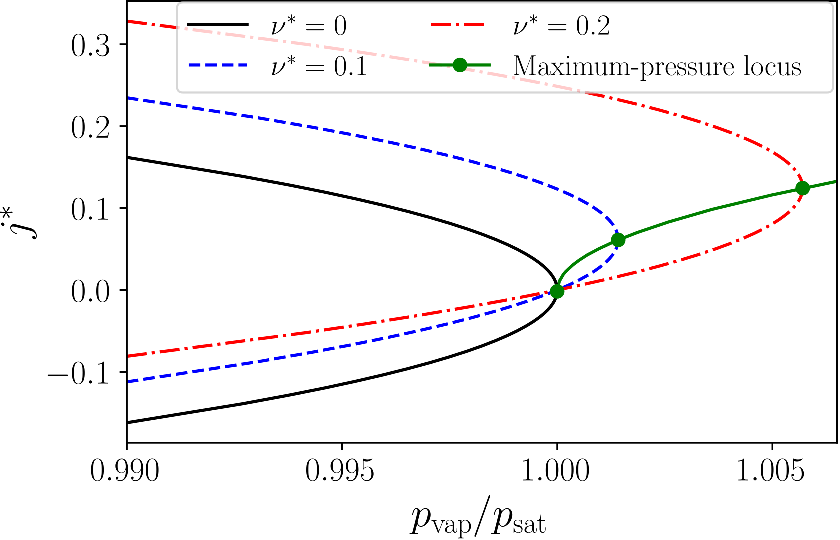}
    \caption{Locally admissible interfacial mass fluxes as functions
    of the adjacent vapor pressure for different dimensionless
    viscosities $\nu^\ast$.}
    \label{fig:Symmetry_Breaking_Analytical}
\end{figure}

The $\nu^\ast=0$ curve in
Fig.~\ref{fig:Symmetry_Breaking_Analytical} summarizes this inviscid
solution structure: evaporation and condensation coexist below
saturation with equal $|j^\ast|$, merge at equilibrium, and have no
stationary continuation above saturation. This result reveals a
counterintuitive feature. Condensation is conventionally produced by
increasing an externally imposed pressure above its equilibrium value
~\cite{qu2025experimental,tinguely2012energy,bao2023impact}, yet the
local inviscid interface admits condensation only as the counterpart
of evaporation below saturation. Before examining how this behavior
is reconciled with the response of the complete system, we first
determine how viscous dissipation modifies the inviscid solution
structure.

For the sign convention used throughout the paper, we adopt the left half of the system in Fig.~\ref{fig:sketch} as the reference domain, with $x$ increasing toward the liquid. Under this
convention, $j>0$ denotes condensation and $j<0$ evaporation
(see Sec.~V\,B of the Supplemental Material).




\emph{Viscous symmetry breaking and persistent multiplicity---}
What happens to the evaporation--condensation symmetry when viscous
dissipation is introduced? To answer this question, we solve the
viscous interfacial problem over a range of local vapor pressures and
viscosities.

The constitutive behavior of viscous stresses within a diffuse
liquid--vapor interface is not addressed here. Instead, we adopt a
simple closure by extending the bulk relation throughout the interfacial region, with constant kinematic viscosity $\nu$. This choice also facilitates a direct implementation in the BGK lattice Boltzmann formulation used below. Further details on the resulting viscous contribution are provided in the Supplemental Material Sec.~II\,C.

Figure~\ref{fig:Symmetry_Breaking_Analytical} shows the resulting
solution families for increasing values of $\nu^\ast$.
Once viscosity is introduced, the inviscid symmetry is lost: the
viscous contribution in Eq.~\eqref{eq:interfacial_chemical} changes
under reversal of the flow, so that $j\rightarrow-j$ no longer leaves
the interfacial problem invariant.

Remarkably, breaking the reversible symmetry does not restore a
unique correspondence between the adjacent vapor state and the
phase-change rate. Instead, the solution structure is reorganized as
\[
\begin{array}{ll}
\nu^\ast>0:\quad
&p_{\mathrm{vap}}<p_{\mathrm{sat}}
\longrightarrow \{E,C\},\\[1mm]
&p_{\mathrm{sat}}<p_{\mathrm{vap}}<p_{\max}
\longrightarrow \{C_1,C_2\},\\[1mm]
&p_{\mathrm{vap}}>p_{\max}
\longrightarrow \varnothing ,
\end{array}
\]
where $E$ and $C$ denote evaporation and condensation solutions,
respectively. Below saturation, evaporation and condensation therefore
remain simultaneously admissible, although they are no longer related
by the inviscid symmetry. Above saturation, viscosity creates a new
solution regime in which two distinct stationary condensation rates
coexist for the same adjacent vapor pressure. The two condensation
branches eventually merge at $p_{\max}$, beyond which no stationary
local interfacial solution is found.




\emph{Dynamical selection of phase-change states—}
If multiple phase-change states are locally admissible, which one does the complete system actually realize? The missing information comes from the surrounding fluid. A condition imposed remotely does not act
directly on the interface, but must propagate through the bulk before establishing the adjacent thermodynamic state. Local admissibility and dynamical accessibility must therefore be considered together.

The stationary analysis above requires an approximately isothermal
interfacial region, but does not require the entire system to evolve
isothermally. We now introduce an additional simplification and consider
isothermal dynamics also in the bulk phases of the vapor--liquid--vapor
configuration shown in Fig.~\ref{fig:sketch}. This allows the propagation
through the compressible vapor to be described by isothermal pressure
waves, while retaining the stationary interfacial structure derived
above. Our purpose at this stage is not to reproduce the full 
evolution of a phase-change process, but to isolate
how a stationary interfacial state is dynamically connected to a
pressure perturbation imposed remotely from the interface.

Starting from equilibrium at $p_{\mathrm b}=p_{\mathrm{sat}}$, we impose
$p_{\mathrm b}=p_{\mathrm{sat}}+\Delta p$. The resulting pressure
disturbance propagates through the vapor and establishes the local state
adjacent to the interface. Crucially, the imposed boundary pressure
$p_{\mathrm b}$ is therefore not the local vapor pressure
$p_{\mathrm{vap}}$ entering the stationary interfacial relations; the
two are connected by the dynamics of the intervening vapor.

A boundary perturbation propagates toward the interface as a shock wave
connecting piecewise-uniform vapor states through the isothermal
Rankine--Hugoniot conditions
~\cite{rankine1870thermodynamic,hugoniot1887memoire},
\begin{subequations}
\label{eq:rankine_hugoniot}
\begin{equation}
j_{\mathrm s}
=
\rho_{\mathrm L}(u_{\mathrm L}-s)
=
\rho_{\mathrm R}(u_{\mathrm R}-s),
\end{equation}
\begin{equation}
p_{\mathrm L}
+\frac{j_{\mathrm s}^2}{\rho_{\mathrm L}}
=
p_{\mathrm R}
+\frac{j_{\mathrm s}^2}{\rho_{\mathrm R}},
\end{equation}
\end{subequations}
where the subscripts $\mathrm L$ and $\mathrm R$ denote the fluid states
to the left and right of the shock, respectively, $j_{\mathrm s}$ is
the mass flux across the shock, and $s$ is the shock velocity. The selection process can then be summarized as
\[
A
\xrightarrow{\mathrm{incoming\ shock}}
B
\xrightarrow{\mathrm{interfacial\ relaxation}}
C ,
\]
where $A$ is the initial equilibrium vapor state, $B$ is established by the boundary-generated wave, and $C$ is the phase-change state reached after interaction with the interface. The selected state must simultaneously belong to the local interfacial solution family and be
dynamically reachable from $B$ through an outgoing shock. In this sense,
\[
C
=
\{\mathrm{admissible}\}
\cap
\{\mathrm{reachable\ from}\ B\}.
\]
The detailed construction of the states $A$, $B$, and $C$ from the
Rankine--Hugoniot and local interfacial relations, together with the
long-domain assumption used to isolate this initial selection process,
is given in Sec.~IV of the Supplemental Material.

This sequence is observed directly in the lattice Boltzmann dynamics. Details of the lattice Boltzmann method, simulation setup, and parameters used for the simulations in the figures are provided in Secs.~V\,A--V\,C of the Supplemental Material, respectively.

\begin{figure}
    \centering
    \includegraphics[width=0.5\textwidth]{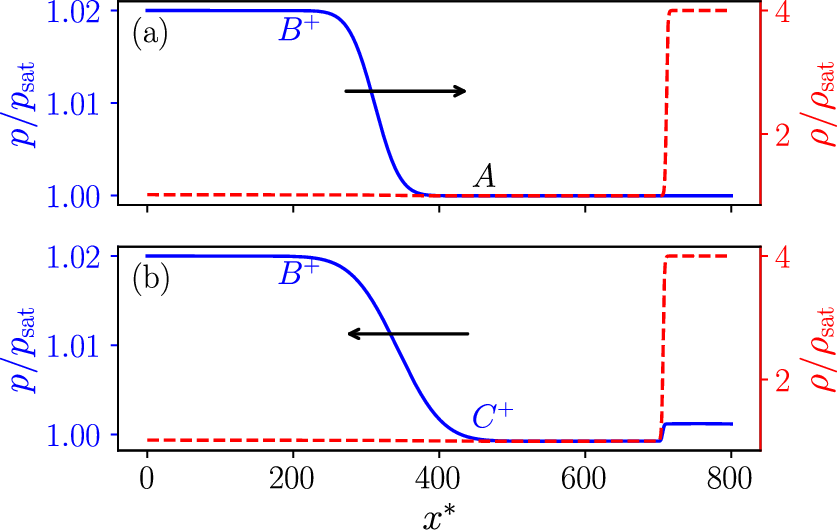}
    \caption{\label{fig:SystemEvolution}
    Lattice Boltzmann snapshots following a positive boundary-pressure
    perturbation. (a) The incoming shock changes the vapor from $A$ to
    $B^{+}$. (b) Interfacial relaxation selects the condensation state
    $C^{+}$ and generates an outgoing shock.
    }
\end{figure}

For a positive pressure perturbation, the incoming shock first establishes $B^{+}$ while the interface remains close to equilibrium
[Fig.~\ref{fig:SystemEvolution}(a)]. Once the shock reaches the interface, condensation develops toward $C^{+}$ and an outgoing shock propagates back through the vapor
[Fig.~\ref{fig:SystemEvolution}(b)]. The simulations therefore realize the same $A\rightarrow B\rightarrow C$ mechanism used in the
analytical construction.

\begin{figure*}
    \centering
    \includegraphics[width=\textwidth]{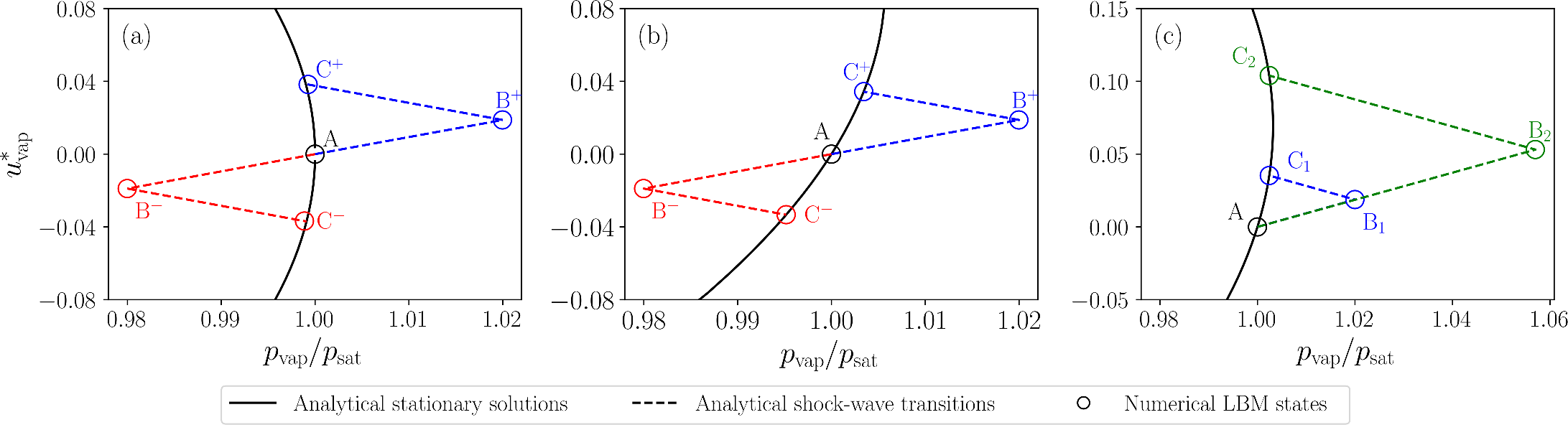}
    \caption{\label{fig:DynamicalSelection}
Dynamical selection of phase-change states.
(a) In the inviscid limit, positive and negative boundary-pressure
perturbations generate
$A\rightarrow B^{+}\rightarrow C^{+}$ and
$A\rightarrow B^{-}\rightarrow C^{-}$, selecting condensation and
evaporation, respectively.
(b) With viscosity, perturbations of equal magnitude and opposite signs
select evaporation and condensation states that are no longer related
by the inviscid symmetry; the condensation state lies above saturation.
(c) Different positive perturbations from the same equilibrium state
select $C_1$ and $C_2$, which have approximately the same local vapor
pressure but different condensation rates.
}
\end{figure*}

The consequence for the inviscid multiplicity is shown in
Fig.~\ref{fig:DynamicalSelection}(a). Although evaporation and
condensation are both locally admissible below saturation, the complete system does not select them arbitrarily. We find
\[
\Delta p_{\mathrm b}>0
\ \longrightarrow\ C,
\qquad
\Delta p_{\mathrm b}<0
\ \longrightarrow\ E .
\]
Increasing the externally imposed pressure therefore selects
condensation, whereas decreasing it selects evaporation, recovering the conventional pressure-driven response. At the same time, the local result derived above remains intact: the condensation state
$C^{+}$ selected by $\Delta p_{\mathrm b}>0$ lies on the
sub-saturation branch,
$p_{\mathrm{vap}}(C^{+})<p_{\mathrm{sat}}$. Thus, the pressure increase that initiates condensation is a property of the remotely imposed condition and need not correspond to a super-saturation vapor pressure adjacent to the interface.

Viscosity produces two distinct changes in the dynamical selection.
First, Fig.~\ref{fig:DynamicalSelection}(b) contrasts directly
with the inviscid result in Fig.~\ref{fig:DynamicalSelection}(a). In the
inviscid limit, boundary-pressure perturbations of equal magnitude and
opposite signs select evaporation and condensation states related by
the reversible symmetry and lying at approximately the same local vapor
pressure. With viscosity, the corresponding dynamical paths no longer
terminate at symmetry-related states. In particular, a positive
boundary-pressure perturbation can select a condensation state above
saturation.

Second, Fig.~\ref{fig:DynamicalSelection}(c) shows that the two
condensation solutions available at approximately the same local vapor
pressure are both dynamically accessible. Starting from the same
equilibrium state, different positive boundary-pressure perturbations
produce
\[
A\rightarrow B_1\rightarrow C_1,
\qquad
A\rightarrow B_2\rightarrow C_2,
\]
with
\[
p_{\mathrm{vap}}(C_1)\simeq p_{\mathrm{vap}}(C_2),
\qquad
j(C_1)\neq j(C_2).
\]
The viscous multiplicity is therefore not merely a set of redundant
local solutions: distinct dynamical paths can select different
condensation rates at essentially the same adjacent vapor state.

The apparent conflict created by the local multiplicity is thus resolved once the interface is coupled to its surroundings. The local interfacial relations determine which phase-change states are admissible, while the bulk dynamics determines which of them is reached from prescribed initial and boundary conditions:
\[
\begin{array}{rcl}
\mathrm{interface} &\longrightarrow& \mathrm{admissibility},\\
\mathrm{bulk\ dynamics} &\longrightarrow& \mathrm{selection}.
\end{array}
\]
A local vapor pressure alone therefore need not determine either the direction or the rate of phase change. The conventional response to an externally imposed pressure perturbation emerges from the coupled interface--bulk dynamics that selects among the locally admissible
states.

Because the lattice Boltzmann simulations evolve the complete
diffuse-interface system without imposing the integrated interfacial
relations or selecting their branches, their agreement with the analytical predictions shows that the admissible states are dynamically realizable, rather than merely formal solutions of the reduced interfacial problem.








\emph{Conclusion—}
We have uncovered a stationary phase-change structure in which
evaporation and condensation emerge as reversible counterparts in the inviscid limit. For $p_{\mathrm{vap}}<p_{\mathrm{sat}}$, every
stationary evaporation state has a corresponding condensation state
with the same adjacent thermodynamic conditions and opposite mass
flux. Viscous dissipation breaks this symmetry but does not
restore uniqueness: evaporation and condensation remain simultaneously
admissible below saturation, while two distinct condensation rates can
coexist above saturation.

The realization of these locally admissible states is determined by
the surrounding dynamics. In the isothermal dynamical problem considered
here, a remotely imposed pressure perturbation propagates through the
compressible vapor and selects condensation for increasing boundary
pressure and evaporation for decreasing pressure. Distinct dynamical
paths can also select different condensation rates at approximately the
same local vapor pressure. Lattice Boltzmann simulations reproduce both
the predicted stationary states and their dynamical selection.

These results show that the local thermodynamic state alone need not
uniquely determine either the direction or the rate of phase change.
Instead, diffuse-interface theory separates the problem into local
interfacial admissibility and dynamical selection by the surrounding
fluid. Whether this structure has a counterpart in molecular and
experimental phase-change dynamics remains an open question.

\emph{Acknowledgment-}
This study was financed, in part, by the São Paulo Research Foundation (FAPESP), Brasil, Process Numbers 2025/11714-1 and 2022/15765-1.



\bibliography{apssamp}

\end{document}